\documentclass[aps,prl,twocolumn,amsfonts,amssymb,amsmath,floats,floatfix,showpacs,preprintnumbers,superscriptaddress]{revtex4-1}

\usepackage{graphicx}
\usepackage{dcolumn}
\usepackage{bm}

\begin{document}

\title{Rashba spin-splitting control at the surface of the topological insulator Bi$_2$Se$_3$}

\author{Z.-H. Zhu}
\affiliation{Department of Physics {\rm {\&}} Astronomy, University of British Columbia, Vancouver, British Columbia V6T\,1Z1, Canada}
\author{G. Levy}
\affiliation{Department of Physics {\rm {\&}} Astronomy, University of British Columbia, Vancouver, British Columbia V6T\,1Z1, Canada}
\author{B. Ludbrook}
\affiliation{Department of Physics {\rm {\&}} Astronomy, University of British Columbia, Vancouver, British Columbia V6T\,1Z1, Canada}
\author{C.N. Veenstra}
\affiliation{Department of Physics {\rm {\&}} Astronomy, University of British Columbia, Vancouver, British Columbia V6T\,1Z1, Canada}
\author{J.A. Rosen}
\affiliation{Department of Physics {\rm {\&}} Astronomy, University of British Columbia, Vancouver, British Columbia V6T\,1Z1, Canada}
\author{R. Comin}
\affiliation{Department of Physics {\rm {\&}} Astronomy, University of British Columbia, Vancouver, British Columbia V6T\,1Z1, Canada}
\author{D.\,Wong}
\affiliation{Department of Physics {\rm {\&}} Astronomy, University of British Columbia, Vancouver, British Columbia V6T\,1Z1, Canada}
\author{\\P. Dosanjh}
\affiliation{Department of Physics {\rm {\&}} Astronomy, University of British Columbia, Vancouver, British Columbia V6T\,1Z1, Canada}
\author{A. Ubaldini}
\affiliation{D\'{e}partment\,de\,Physique\,de\,la\,Mati\`{e}re\,Condens\'{e}e, Universit\'{e} de Gen\`{e}ve, CH-1211 Gen\`{e}ve 4, Switzerland}
\author{P. Syers}
\affiliation{CNAM, Department of Physics, University of Maryland, College Park, Maryland 20742, USA}
\author{N.P. Butch}
\affiliation{CNAM, Department of Physics, University of Maryland, College Park, Maryland 20742, USA}
\author{J. Paglione}
\affiliation{CNAM, Department of Physics, University of Maryland, College Park, Maryland 20742, USA}
\author{I.S. Elfimov}
\affiliation{Department of Physics {\rm {\&}} Astronomy, University of British Columbia, Vancouver, British Columbia V6T\,1Z1, Canada}
\affiliation{Quantum Matter Institute, University of British Columbia, Vancouver, British Columbia V6T\,1Z4, Canada}
\author{A. Damascelli}
\email{damascelli@physics.ubc.ca}
\affiliation{Department of Physics {\rm {\&}} Astronomy, University of British Columbia, Vancouver, British Columbia V6T\,1Z1, Canada}
\affiliation{Quantum Matter Institute, University of British Columbia, Vancouver, British Columbia V6T\,1Z4, Canada}

\date{\today}

\begin{abstract}
The electronic structure of Bi$_2$Se$_3$ is studied by angle-resolved photoemission and density functional theory. We show that the instability of the surface electronic properties, observed even in ultra-high-vacuum conditions, can be overcome via in-situ potassium deposition. In addition to accurately setting the carrier concentration, new Rashba-like spin-polarized states are induced, with a tunable, reversible, and highly stable spin splitting. Ab-initio slab calculations reveal that these Rashba state are derived from the 5QL quantum-well states. While the K-induced potential gradient enhances the spin splitting, this might be already present for pristine surfaces due to the symmetry breaking of the vacuum-solid interface.
\end{abstract}

\pacs{71.20.-b, 71.10.Pm, 73.20.At, 73.22.Gk}

\maketitle

Topological insulators, with a gapless topological surface state (TSS) located in a large bulk bandgap, define a new quantum phase of matter \cite{Fu:2007, Qi:2008, Moore:2010, Hasan:2010PRM}. Their uniqueness, and their strong application potential in quantum electronic devices, stem from the TSS combination of spin polarization and protection from backscattering \cite{Bernevig:HgTe, Roushan:scattering}. Bi$_2$Se$_3$ is a three dimensional topological insulator, as theoretically proposed \cite{Zhang:2009BeSe} and experimentally verified by angle-resolved photoemission spectroscopy (ARPES) and other surface sensitive techniques \cite{Xia:2009BiSe, Chen:2010BiSe, Cheng:2010STM}. Unfortunately, despite great effort in controlling the Se stoichiometry and with it the bulk carrier concentration \cite{JP}, unintentional and uncontrolled doping seems to lead to a bulk conductivity that masks the surface electronic properties \cite{Hirahara:2010}. 
\begin{figure}[b!]
\includegraphics[width=0.92\linewidth]{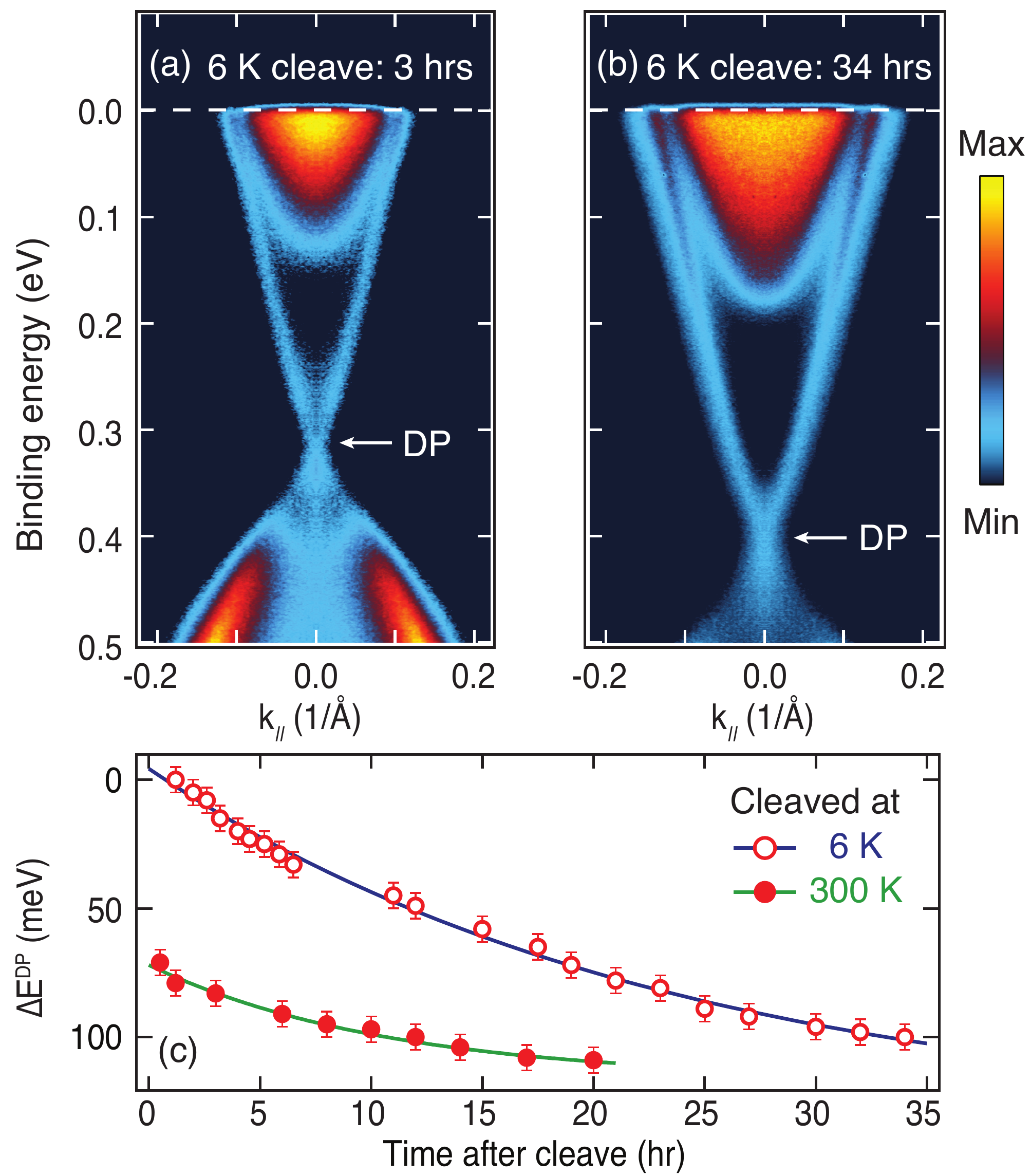}
\caption{\label{fig:time} (color online). (a,b) Time evolution of the ARPES dispersion of Bi$_2$Se$_3$ at 5$\times$10$^{-11}$\,torr and $T\!=\!6$\,K: (a) 3 hours after cleaving; (b) 34 hours after cleaving. (c) Exponential fit of the Dirac point (DP) binding energy versus time for 6 and 300\,K cleaves (both measured at 6\,K); saturation at $\Delta E^{DP}\!\simeq\!133$ and 116\,meV is reached in 46 and 22 hours, respectively.}
\end{figure}
ARPES studies also have shown that cleaved sample surfaces and subsurfaces become progressively more electron doped -- even in ultra-high vacuum conditions -- by either gas adsorption, or formation/migration of defects and vacancies \cite{Park:2010,Hsieh:2009}. Lastly, the TSS might be completely destroyed when exposed to air, hindering most attempts of material processing and characterization, as well as device fabrication.

Developing new approaches to stabilize and control the surface of these systems is arguably the most critical step towards the exploitation of their topological properties. Some success has been obtained in inducing electron and hole surface doping by a combination of in-situ processing, such as material evaporation and radiation exposure \cite{Hsieh:2009BiSe, Xu:doping}. 
\begin{figure*}[t!]
\includegraphics[width=1\linewidth]{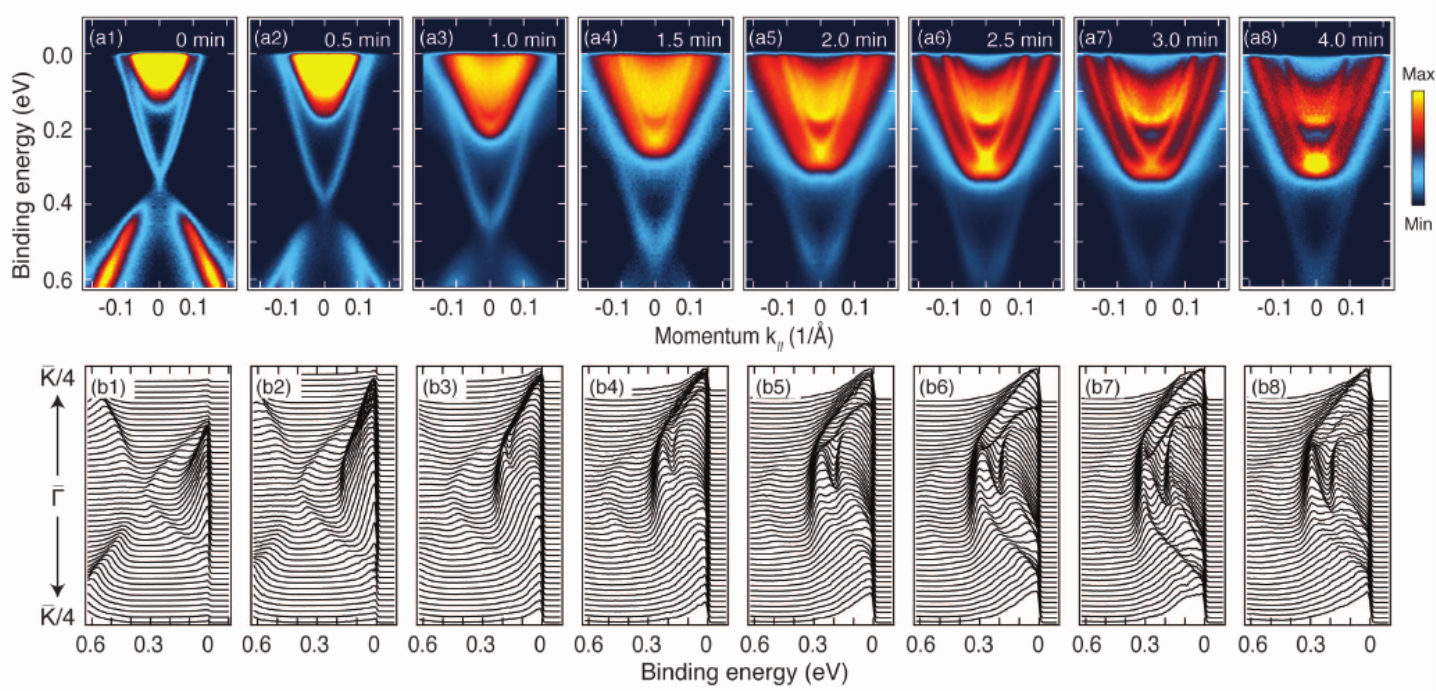}
\caption{\label{fig:kevap} (color online).\,Evolution of the Bi$_2$Se$_3$ $\bar{\Gamma}\!-\!\bar{K}$ electronic dispersion upon subsequent 0.5\,min K-evaporation steps:\,(a1-a8) ARPES image plots, (b1-b8) corresponding energy distribution curves (EDCs). The sample was kept at 5$\times$10$^{-11}$\,torr and 6\,K.} 
\end{figure*}
The same TSS has also been fabricated on nanoribbons, which have large surface-to-volume ratio \cite{Kong:2010}. From a different perspective, carefully doped topological insulators can provide a platform to study the interplay between TSS and bulk electron dynamics, which has important implications for TSS control and exploring topological superconductivity \cite{Fu:SCTI}. 

In this paper, we present a systematic ARPES study of the evolution of the surface electronic structure of Bi$_2$Se$_3$ as a function of time and in-situ potassium evaporation. The deposition of submonolayers of potassium allows us to stabilize the otherwise continually evolving surface carrier concentration. It also leads to a more uniform surface electronic structure, in which well-defined Rashba-like states emerge from the continuum of parabolic-like states that characterizes the as-cleaved, disordered surfaces. This approach provides a precise handle on the surface doping, and also allows tuning the spin splitting of the Rashba-like states. Our density functional theory (DFT) slab calculations reveal that the new spin-split states originate from the  bulk-like quantum-well (QW) states of a 5-quintuple-layer (5QL) slab, as a consequence of the K-enhanced inversion symmetry breaking already present for the pristine surface of Bi$_2$Se$_3$.

ARPES measurements were performed at UBC with 21.2\,eV linearly polarized photons on an ARPES spectrometer equipped with a SPECS Phoibos 150 hemispherical analyzer and UVS300 monochromatized gas discharge lamp. Energy and angular resolution were set to 10\,meV and 
$\pm$0.1$^{\circ}$. Bi$_2$Se$_3$ single crystals, grown from the melt (with carrier density $n\!\simeq\!1.24\!\times\!10^{19}$\,cm$^{-3}$ \cite{JP}) and by floating zone, were aligned by Laue diffraction and cleaved and measured at pressures better than 5$\times$10$^{-11}$\,torr and 6\,K, unless otherwise specified. No difference was observed for samples grown with different methods. Potassium was evaporated at 6K, with a 6.2\,A evaporation current for 30 second intervals \cite{Suman,David}. DFT calculations were performed using the linearized augmented-plane-wave method in the WIEN2K package \cite{wien2k}, with structural parameters from Ref.\,\onlinecite{Zhang:2009BeSe}. We considered stoichiometric slabs terminated by a Se layer on both sides, representing natural cleavage planes within this material. Spin-orbit interaction is included as a second variational step using scalar-relativistic eigenfunctions as a basis \cite{wien2k}; exchange and correlation effects are treated within the generalized gradient approximation \cite{GGA}.

\begin{figure}[t!]
\includegraphics[width=1\linewidth]{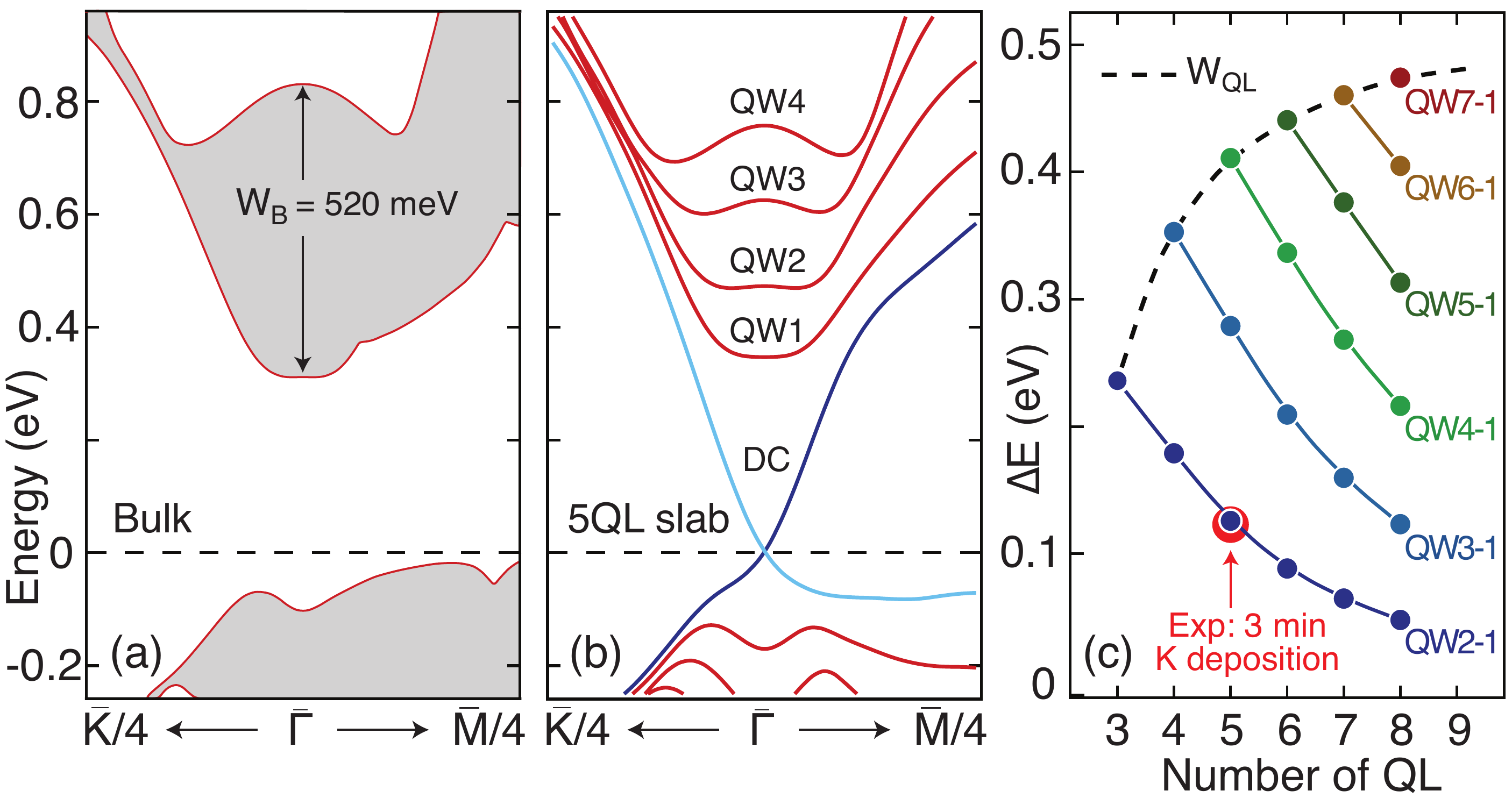}
\caption{\label{fig:LDA} (color online). DFT results for (a) $k_z-$projected bulk and (b) 5QL slab of Bi$_2$Se$_3$ ($E_F$ is at energy 0). (c) Energy difference of the quantum-well (QW) slab states with respect to QW1, calculated for various thicknesses. The QW number increases with QL number, defining a bandwidth $W_{QL}$ that asymptotically approaches be fully $k_z-$projected bulk $W_B\!=\!520$\,meV. The 3 minute K-deposition RB2-RB1 splitting of  $123\!\pm\!6$\,meV  is accurately reproduced by the 5QL results.}
\end{figure}

The time evolution of the as-cleaved Bi$_2$Se$_3$ surface is shown in Fig.\ref{fig:time}. As typically observed by ARPES, and contrary to what is predicted by DFT for fully stoichiometric Bi$_2$Se$_3$ (Fig.\,\ref{fig:LDA}), even immediately after a 6\,K cleave the Fermi level is not in the bulk gap; instead it crosses both TSS and parabolic continuum of bulk-like states. The pronounced time dependence of the data is exemplified by the variation of the Dirac point (DP) binding energy ($\Delta E^{DP}$), which increases from $\sim\!300$ to 400\,meV over 34 hours at 5$\times$10$^{-11}$\,torr and 6\,K [Fig.\,\ref{fig:time}(c)]. An exponential fit of $\Delta E^{DP}$ versus time indicates that the saturation value $E^{DP}\!\simeq\!433$\,meV is reached 46 hours after cleaving. At variance with the time dependence of the TSS, the bottom of the parabolic continuum shifts down by only 30\,meV in 34 hours, which provides evidence against the pure surface nature of the continuum. One should note that the pristine position of DP depends also on the cleave temperature: on a sample cleaved at 300\,K we found a 70\,meV deeper starting position for the DP, although the saturation value is approximately the same as that of the 6\,K cleave [Fig.\,\ref{fig:time}(c)]. 
 
\begin{figure}[t!]
\includegraphics[width=0.98\linewidth]{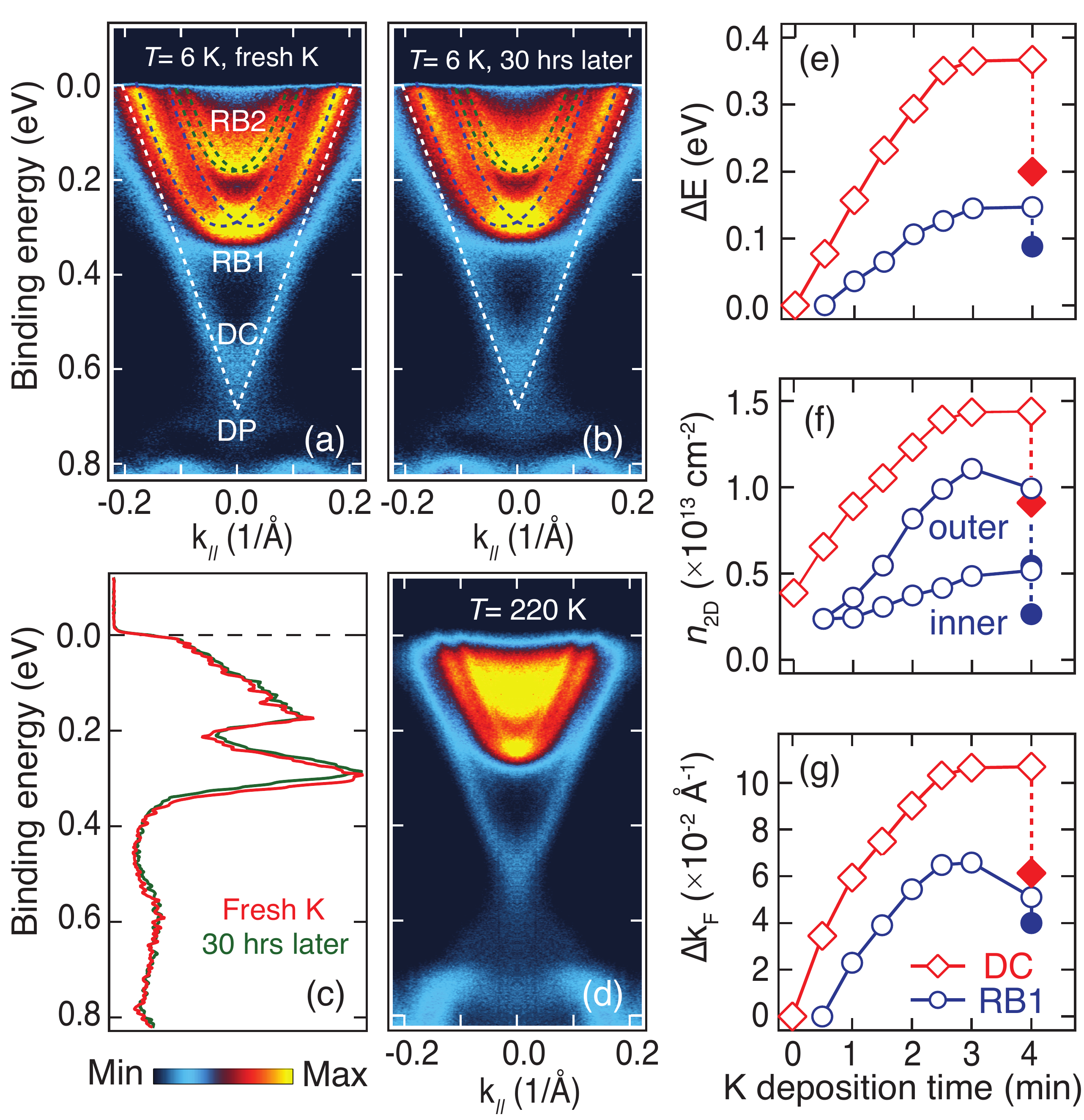}
\caption{\label{fig:summary} (color online). ARPES $\bar{\Gamma}\!-\!\bar{K}$ band dispersion from Bi$_2$Se$_3$ taken (a) immediately after a 3\,min K evaporation, and (b) 30 hours later (the sample was kept at 5$\times$10$^{-11}$\,torr and 6\,K the whole time). As also emphasized by the comparison of the corresponding $\bar{\Gamma}$ point EDCs in (c), the evaporated surface is highly stable. (d) Band dispersion measured at 220\,K after a slow 36 hour warming up on a sample initially K evaporated for 4\,min at 6\,K; the comparison with the data in Fig.\,\ref{fig:kevap}(a8) reveals the suppression of the K-induced carrier doping. (e-g) Evolution vs. K-evaporation time of: (e) binding energy variation for DP  ($\Delta E^{DP}$) and bottom of RB1 ($\Delta E^{RB1}$), as defined in (a); (f) sheet carrier density for DC ($n_{2D}^{DC}$) and RB1 ($n_{2D}^{RB1}$); (g) variation of the DC Fermi wavevector ($\Delta k_F^{DC}$) and of the Rashba band splitting at $E_F$ ($\Delta k_F^{RB1}$). Empty symbols in (e-g) are for $T\!=\!6$\,K and filled ones for $T\!=\!220$\,K.}
\end{figure}

In our ARPES study, the surface time evolution resulted only in the deepening of Dirac cone (DC) and bulk continuum, as a consequence of the sample gaining electrons. Other effects, such as the reported appearance of a 2-dimensional electron gas (2DEG), were not observed \cite{Bianchi:20102DEG}. More substantial changes are induced by the in-situ evaporation of potassium on the cleaved surfaces, also performed at 6\,K to guarantee the highest stability. As a function of K-deposition time, three stages can be identified: {\it Stage I} -- for moderate K deposition [up to 1 minute, Fig.\,\ref{fig:kevap}(ab1-ab3)], the DP moves to higher binding energy by electron doping and a sharper parabolic state appears at the edge of the bulk continuum, reminiscent of the proposed 2DEG \cite{Bianchi:20102DEG}. {\it Stage II}: for intermediate K deposition [from 1 to 3 minutes, Fig.\,\ref{fig:kevap}(ab4-ab7)], as the electron doping further increases, the single parabola splits into a first pair -- and then a second one also appears -- of sharp parabolic states with an equal and opposite momentum shift away from the $\bar{\Gamma}$ point, as in a Rashba type \cite{Rashba} splitting [these states are labeled RB1 and RB2 in Fig.\,\ref{fig:summary}(a)]. Interestingly, the appearance of the sharp RB1 and RB2 features is accompanied by a suppression of the bulk-like continuum. This emergence of a coherent quasiparticle dispersion from a continuum of incoherent spectral weight indicates that the evaporation of potassium leads to a progressively more uniform surface and subsurface structure. {\it Stage III}:  for heavy K deposition [beyond 3 minutes, Fig.\,\ref{fig:kevap}(ab8)], the bottom of RB1 and RB2 as well as $E^{DP}$ are not changing, indicating that the sample cannot be doped any further. The only noticeable effect is a small decrease of spin splitting for RB1 and conversely an increase for RB2, perhaps stemming from a change in hybridization between the two Rashba pairs. As a last remark, during the entire K-deposition process the band velocity of the TSS close to the DP is $3.2\pm0.3$\,eV$\mathrm{\AA}$, consistent with previous reports \cite{Kuroda:2010Hex}. 
 
Before analyzing quantitatively the evolution of the various states upon K deposition, we address the question of the stability of this new surface versus time and temperature cycling. In Fig.\,\ref{fig:summary}(a-c) we compare the ARPES data from a 3 minute K-evaporated surface, as measured right after deposition and 30 hours later (during which the sample was kept at 6\,K). Other than a smaller than 10\,meV shift of the bottom of RB1 [Fig.\,\ref{fig:summary}(c)], all spectral features including the TSS have remained exactly the same over the 30 hour interval. This is a remarkable stability, especially when compared to the 365\,meV shift induced by the initial K deposition [Fig.\,\ref{fig:summary}(e)], and to the more than 100\,meV shift observed versus time without any active surface processing [Fig.\,\ref{fig:time}(c)]. This approach might provide a new path to overcome the general instability and self-doping problem of the surface of  Bi$_2$Se$_3$, which represents one of the major shortcomings towards the fabrication of topological devices.  Temperature effects were studied by slowly warming up the sample, in which case K atoms diffuse and eventually leave the surface, reverting the material back to an earlier stage with lower K coverage. Indeed, as one can see by comparing Fig.\,\ref{fig:summary}(d) to Fig.\,\ref{fig:kevap}(a4), a sample initially K evaporated for 4 minutes at 6\,K, and then measured at 220\,K after a gradual 36 hour warming up, exhibits ARPES features similar to those obtained directly after a 1.5 minute K deposition at 6\,K. This implies that K deposition on Bi$_2$Se$_3$ is also reversible, making it possible to fine tune surface doping, position of the DP, and Rashba spin splitting.

We summarize in Fig.\,\ref{fig:summary}(e-g) the K-evaporation evolution of various parameters characterizing the $\bar{\Gamma}\!-\!\bar{K}$ dispersion of DC and Rashba states (empty symbols identify 6\,K data, and the filled ones 220\,K data). As evident in Fig.\,\ref{fig:summary}(e) from the variation of $E^{DP}$ and bottom of RB1, the highest possible doping level is achieved $\sim$3 minutes into the K deposition, corresponding to $\Delta E^{DP}\!\simeq\!365$\,meV and $\Delta E^{RB1}\!\simeq\!150$\,meV. 
The K-induced change in surface electron density for the various states can be estimated from the relation $n_{2D}\!=\!A_{FS}/A_{BZ}\,A_{UC}$ between the area of Fermi surface, Brillouin zone, and unit cell, without accounting for spin degeneracy given that  all relevant states are spin split. Because at these electron fillings all FS's are hexagonal, this reduces to $n_{2D}\!=\!k_F^2/2\sqrt3\pi^2$, where $k_F$ is the Fermi wavevector along the $\bar{\Gamma}\!-\!\bar{K}$ direction of the BZ (as in Fig.\,\ref{fig:kevap}). 
After 3 minute K evaporation the total sheet carrier density is $n_{2D}^{tot}\!\simeq\!3.64\times10^{13}$\,cm$^{-2}$ (0.162\,electron/BZ), corresponding to the sum of the contributions from DC, and inner-and-outer RB1 and RB2 (1.43, 1.60, and $0.61\times10^{13}$\,cm$^{-2}$ respectively). This value is to be compared to $n_{2D}^{tot}\!\simeq\!3.87\times10^{12}$\,cm$^{-2}$ before K deposition (0.017\,electron/BZ), which however only accounts for the DC, given the impossibility of estimating the contribution from the parabolic continuum.

As a last point, from the $\Delta k_{F}$ data presented in Fig.\,\ref{fig:summary}(g), and the dispersion of spin-split Rashba bands:
\begin{equation}
E^{\pm}(k_{\|}) = E_{\bar{\Gamma}} + \frac{\hbar^2k_{\|}^2}{2m^{\star}}\pm\alpha_Rk_{\|},
\label{eq:one}
\end{equation}

\noindent we can estimate the Rashba parameter $\alpha_R\!=\!\hbar^2 \Delta k_{\|}^{\pm}/2m^{\star}$ for RB1. The latter, which depends both on the value of spin-orbit coupling (SOC) and the gradient of the potential $\partial V/\partial z$ \cite{Nagano}, reflects the size of the spin splitting in momentum space and is here controlled directly by the amount of K deposited on the as-cleaved surfaces. The largest RB1 splitting is observed after 2.5-3 minute K evaporation and is anisotropic: $\Delta k_{F}\!\simeq\!0.066\mathrm{\AA^{-1}}$ along $\bar{\Gamma}$-$\bar{K}$, and $0.080\mathrm{\AA^{-1}}$ along $\bar{\Gamma}$-$\bar{M}$. Fitting the RB1 dispersion along $\bar{\Gamma}$-$\bar{K}$ to Eq.\,\ref{eq:one}, we obtain $E_{\bar{\Gamma}}\!=\!280\!\pm\!10$\,meV, $m^{\star}\!=\!0.28\!\pm\!0.02m_e$, and $\alpha_R\!=\!0.79\!\pm\!0.03$\,eV\AA. This value is more than twice the Rashba splitting of the Au(111) surface state ($\alpha_R\!\simeq\!0.33$\,eV\AA), and also larger than the one of the Bi(111) surface state ($\alpha_R\!\simeq\!0.56$\,eV\AA) \cite{Ast:Bi}. 

DFT calculations for bulk Bi$_2$Se$_3$, as well as slabs with varying number of QL's (Fig.\,\ref{fig:LDA}), provide a detailed explanation for our observations and some interesting insights. Each QL consists of 2 Bi and 3 Se layers alternating along the c axis, with one Se layer in the middle of the QL and the other two on either side. This forms a non-polar structure with a natural cleavage plane between two adjacent Se layers belonging to different QL's. As shown in Fig.\,\ref{fig:LDA}(b) for the particular case of a 5QL slab, in addition to the TSS-DC there are 4 QW states, for a total of 5 states matching the number of QL's. As evidenced by the comparison with the fully $k_z-$projected bulk results in Fig.\,\ref{fig:LDA}(a), where the TSS is missing due to the absence of the surface, the slab QW states exhibit the same character and energy as the Bi-Se conduction band. However, they are discrete in nature due to quantum confinement, and span a narrower energy range than the corresponding bulk bandwidth $W_B\!=\!520$\,meV. The effective slab bandwidth $W_{QL}$, defined as the energy difference between top and bottom QW states, is asymptotically approaching the bulk $W_B$ value [Fig.\,\ref{fig:LDA}(c)]; for a proper correspondence with the bulk electronic structure a rather large number of QL's is needed (i.e., more than 10 QL). Interestingly, the splitting between the DP and the different QW states is extremely sensitive to the number of QL's. For 5QL we obtain 346\,meV QW1-DP and 126\,meV QW2-QW1 splittings [Fig.\,\ref{fig:LDA}(c)], which closely match the 3 minute K-deposition values $380\!\pm\!50$\,meV and $123\!\pm\!6$\,meV for RB1-DP and RB2-RB1, respectively [as defined from the EDC's at the $\bar{\Gamma}$ point in Fig.\,\ref{fig:summary}(a)]. 

This analysis leads to several important conclusions: {\it (i)} The RB1 and RB2 states that emerge from the parabolic continuum are of the same Bi and Se character than those obtained, in the same energy range, in bulk Bi$_2$Se$_3$ calculations. However, because of the observed lack of $k_z$ dispersion \cite{Bianchi:20102DEG}, and the almost perfect match with QW1 and QW2 energy positions for 5QL [as also emphasized by the comparison between calculation and 3 minute K-deposition datapoint in Fig.\,\ref{fig:LDA}(c)], these states should be more appropriately thought as the quantum-confined analog of those bulk states, associated with a band-bending over a 5QL subsurface region (47.7\,\AA). While this region is somewhat disordered -- either in its depth and/or carrier concentration -- on the as-cleaved surfaces, it becomes progressively better defined upon K evaporation. {\it (ii)} Potassium, in addition to doping carriers, also induces a change in $\partial V/\partial z$, which in turn provides a very direct control knob on both band-bending depth and spin splitting of the Rashba states. {\it (iii)} In light of the extent of the subsurface band-bending region, these quantum-confined states should affect more than just surface sensitive experiments. For instance, Rashba spin-split states might have to be accounted for in the interpretation of transport data even from pristine surfaces, although with a much smaller splitting induced solely by the symmetry breaking vacuum-solid interface.

We thank E. Giannini, A.F. Morpurgo, M. Franz, H. Guo, A.N. Yaresko, M.W. Haverkort, and G.A. Sawatzky for discussions, D. Fournier, D. Schneider, H.R. Davis, M. O'Keane, and W.N. Hardy for technical assistance. This study was supported by the Killam Program (A.D.), Sloan Foundation (A.D.), CRC Program (A.D.), Stacie NSERC Fellowship Program (A.D.), NSERC, CFI, CIFAR Quantum Materials, and BCSI. Work at the University of Maryland was supported by NSF-MRSEC (DMR-0520471) and DARPA-MTO award (N66001-09-c-2067).

\providecommand{\noopsort}[1]{}\providecommand{\singleletter}[1]{#1}%

\end{document}